\documentclass[final,3p,times,numbers]{elsarticle}

\usepackage{natbib}
\usepackage{graphicx}
\usepackage{amssymb,amsmath}
\usepackage[bookmarks={false}]{hyperref}
\usepackage{verbatim}
\usepackage{xcolor}

\journal{Nuclear Physics A}
\begin{document}

\begin{frontmatter}
\center{Nucl.Phys. A1074 (2026) 123462}
\vspace*{3.cm}
\title{Resonances extracted in truncated partial-wave analysis are \\ effective mixtures of angular momenta \\ (Possible implications for H\"ohler's clustering)}

\author[add1,add2]{Alfred \v{S}varc}
\ead{alfred.svarc@gmail.com}

\affiliation[add1]{organization={Rudjer Bo\v{s}kovi\'{c} Institute},%
    addressline={Bijeni\v{c}ka cesta 54},
    city={Zagreb},
    postcode={10000},
    country={Croatia}}

\affiliation[add2]{organization={Tesla Biotech d.o.o},%
    addressline={Mandlova 7},
    city={Zagreb},
    postcode={10000},
    country={Croatia}}

\begin{abstract}
In a truncated partial-wave analysis, one does not fit the amplitudes directly but rather the observables, which are bilinear functions of those amplitudes. For the angle-dependent observables from which one actually infers partial-wave content, truncation does more than simply restrict the analysis to resonances appearing in the retained lower partial waves. Because the observables are bilinear in the amplitudes, the extracted lower partial waves are determined by the full coupled interference structure and hence by a coupled nonlinear fit. As a result, truncation does not merely admix higher-wave contributions into the lower sector: it reshuffles the pole-bearing content among partial waves, including the nominally retained lower ones, so that a resonance contribution associated with one exact angular-momentum sector can reappear in several extracted partial waves and therefore lose a unique angular-momentum assignment. The extracted coefficients are thus generally not straightforward projections of the coefficients from the exact infinite partial-wave expansion, but effective mixtures of contributions originating from several angular-momentum sectors.

We illustrate this explicitly with a minimal scalar toy model, in which a Hermitian bilinear represented by a Legendre series truncated at order 2 is fitted by another series truncated at order 1. Even in this most basic setting, the fitted low-order coefficients turn out to depend on bilinear combinations that involve the higher-order pieces of the original amplitude. Consequently, resonance-related quantities obtained from such a truncated analysis should, in general, not be interpreted as resonances with definite angular momentum.

We then explore a possible phenomenological consequence of this mechanism for H\"ohler's observation that resonance poles assigned to different partial waves in $\pi N$ scattering tend to cluster near a few common points in the complex energy plane. If the pole-bearing quantities extracted in a truncated analysis are effective mixtures of contributions from several angular-momentum sectors, then the inferred resonance spectrum can naturally exhibit cross-wave correlations. In that sense, truncation provides a plausible contribution to H\"ohler-type clustering.

\end{abstract}

\begin{keyword}
Legendre polynomials \sep bilinear fitting \sep nonlinear least squares \sep truncation effects \sep H\"ohler clustering \sep partial-wave mixing
\end{keyword}

\end{frontmatter}

\section{Introduction}

Partial-wave methods are among the standard tools for extracting baryon-resonance information from hadronic
and electromagnetic reactions. In single-channel pseudoscalar-meson photoproduction, for example, the reaction is
described by four complex amplitudes and sixteen polarization observables, which has motivated an extensive literature
on complete experiments, amplitude ambiguities, and multipole extraction~\cite{Barker1975,Chiang1997,Sandorfi2011}. In practice, however, one
usually does not attempt a full amplitude reconstruction at fixed angle, but rather a fixed-energy determination of a finite set
of multipoles. This is the domain of truncated partial-wave analysis (TPWA), in which the multipole expansion
is cut off at some maximal orbital angular momentum $\ell_{\max}$ and the remaining parameters are fitted directly to angular
distributions and polarization observables~\cite{Omelaenko1981,Grushin1989,Wunderlich2014,Workman2017,WunderlichThesis}. Another line of work concerns ambiguities and numerical
stability in truncated analyses, including the relation between discrete ambiguities of TPWA and the continuum ambiguities
of the full amplitude~\cite{WunderlichThesis,Wunderlich2017Amb,Svarc2018}. Closely related is the use of Legendre moments as a compact diagnostic of
partial-wave content~\cite{WunderlichMoments2017,WunderlichMoments2020}. More recently, Bayesian TPWA studies have investigated in parallel several truncation
orders and the corresponding posterior and ambiguity structure~\cite{Kroenert2024}.

A central point of the present work is that, in TPWA and related analyses, one does not fit the amplitudes themselves.
One fits observables, and the observables from which partial-wave content is actually inferred are angle-dependent bilinear
functionals of the amplitudes. Truncation therefore does more than simply discard higher partial waves. Because the
observables are bilinear in the amplitudes, the extracted lower partial waves are determined by the full coupled interference
structure and hence by a coupled nonlinear fit rather than by direct projection of the exact ones. As a result, the extracted
quantities are generally not coefficient-by-coefficient projections of the corresponding quantities of the full non-truncated
problem, but effective mixtures of contributions originating from several angular-momentum sectors.

This statement is stronger than the familiar observation that higher partial waves can influence lower Legendre moments
in the partial-wave decomposition of observables. The issue is not only that omitted partial waves generate corrections.
Because the fitted observables are bilinear, changing the truncation order changes the algebraic structure of the fit itself.
Truncation does not merely admix higher-wave contributions into the lower sector: it can reshuffle the pole-bearing
content among partial waves, including the nominally retained lower ones, so that a resonance contribution associated
with one exact angular-momentum sector can reappear in several extracted partial waves and therefore lose a unique
angular-momentum assignment. In that sense, varying $\ell_{\max}$ does not simply refine a fixed solution; it changes which
combinations of pole-bearing contributions from the exact partial-wave decomposition are inferred from the data in each
truncated partial wave. The pole content of the exact and truncated partial-wave decompositions is therefore not the same.

This observation has direct implications for resonance extraction. Since the exact partial-wave coefficients may contain
poles in the complex energy plane, the truncation-induced mixing discussed here is already a mixing of pole-bearing
quantities. The resulting resonance content inferred from a truncated fit should therefore not, in general, be identified
directly with the true resonance content of the exact non-truncated problem. Rather, truncated analyses determine
effective resonance-carrying structures generated by the restricted bilinear fit. To our knowledge, this specific point
has not been formulated explicitly in the previous TPWA literature: while truncated analyses routinely assign resonances
to partial waves, the possibility that truncation itself may destroy a unique angular-momentum assignment of the
extracted resonance content has not been isolated as a separate issue.

A particularly suggestive phenomenological example is provided by a classic observation of H\"ohler in $\pi N$ scattering.
In his analysis of partial waves, H\"ohler noted that resonance poles assigned to different angular momenta and parities
often appear to bunch together near a few common complex energies, and he raised the possibility that the observed
spectrum might be describable in terms of only a few poles of the invariant amplitudes rather than many independent
poles in separate partial waves~\cite{Hoehler1983}. This phenomenon, which we refer to as H\"ohler's clustering, is conceptually
striking precisely because partial waves are normally expected to carry definite angular momentum. The appearance of
closely grouped poles across different waves therefore raises a basic question: can such clustering arise, at least in part,
as an effective consequence of the truncation procedure used to extract partial-wave information in practice?

The problem was discussed explicitly by Kirchbach and by Losanow and Kirchbach, who treated the H\"ohler clusters
as a genuine structural feature of the baryon spectrum~\cite{Kirchbach1997,LosanowKirchbach1998}. In present Particle Data Group compilations,
however, the situation is less visually transparent than in the older Karlsruhe--Helsinki era, because current resonance
listings combine information from a broader set of hadronic and photoproduction analyses and retain only the more
firmly established states in the summary tables~\cite{PDG2024}. At the same time, the PDG emphasizes that reliable resonance
claims require sufficiently complete partial-wave bases and explicit tests of the effect of higher partial waves~\cite{PDG2024}.
This makes the possible role of truncation especially relevant, even if H\"ohler-type clustering is regarded only as one
particularly visible manifestation of a more general problem.

The purpose of this paper is to formulate the truncation mechanism explicitly in the simplest possible setting and
to explore one of its possible phenomenological consequences. We begin by formulating the bilinear truncation problem
for a scalar partial-wave expansion and by demonstrating the mechanism explicitly in a minimal toy model, where
a Hermitian bilinear constructed from a Legendre expansion truncated at order 2 is approximated by one truncated
at order 1. After that, we return to H\"ohler's clustering and discuss how truncation-induced mixing of pole-bearing quantities
can provide a natural source of cross-wave correlations in extracted resonance content. Although the explicit
derivation is given for scalar scattering, the algebraic mechanism is general and applies equally to Legendre-moment
analyses and to TPWA studies of photoproduction observables.

\section{Fitting observables with truncated bilinears}

We now formulate the central mechanism of this paper in the simplest possible setting. To keep the discussion
transparent, we consider elastic $2 \to 2$ scattering of scalar particles. A general scalar scattering process is described
by the reaction amplitude
\begin{equation}
f(W, x) = \sum_{l=0}^{\infty} a_l(W) P_l(x), \label{eq:pwexp}
\end{equation}
where the coefficients $a_l(W)$ may contain poles associated with definite angular momentum $l$, and $P_l(x)$ are
the usual Legendre polynomials. In the exact infinite problem this decomposition is unique, and each partial wave
carries definite angular momentum. In practice, however, one never determines the amplitude directly. One fits observables, and the angle-dependent observables from which partial-wave content is inferred are bilinear in the amplitudes. Truncation therefore does more than simply discard higher partial waves. Once a bilinear quantity of the form $f(W,x)\overline{f(W,x)}$ is approximated by bilinears constructed from a truncated amplitude, the extracted lower partial waves are no longer obtained by direct projection of the exact ones. Instead, they are determined by the full coupled bilinear structure of the fit and therefore become effective mixtures of contributions originating from several angular-momentum sectors.

The essential point is thus not merely that omitted higher partial waves produce corrections. The point is that truncation changes the map between the exact amplitude and the extracted truncated one. In particular, the pole-bearing content associated with one exact angular-momentum sector can be redistributed among several extracted partial waves, including the nominally retained lower ones. This is the mechanism that we now formulate explicitly.

\subsection{Formulation of the bilinear problem}

Since working with an infinite number of terms is impractical, we consider a simplified setting in which a Hermitian
bilinear of Legendre rank $N$ is approximated by Hermitian bilinears of rank $M$ with $M < N$:
\begin{equation}
f(W, x) = \sum_{l=0}^{N} a_l(W) P_l(x), \qquad
g(W, x) = \sum_{m=0}^{M} b_m(W) P_m(x), \qquad M < N,
\end{equation}
with $a_l(W), b_m(W) \in \mathbb{C}$. The corresponding Hermitian bilinears are
\begin{equation}
B_f(W, x) = f(W, x)\overline{f(W, x)}, \qquad
B_g(W, x) = g(W, x)\overline{g(W, x)}.
\end{equation}

Now observe that both bilinears are initially given as products of Legendre polynomials, and this is not an orthogonal basis. The problem is therefore qualitatively different from the familiar truncation of partial-wave amplitudes themselves, where the expansion is already written in an orthogonal basis. In that simpler case, orthogonality implies that the best lower-order approximation is obtained by keeping the lower coefficients unchanged, while the higher partial waves are simply omitted. For the bilinear problem this is no longer true. In order to compare $B_f(W,x)$ and $B_g(W,x)$, both quantities must first be transformed into the orthogonal basis of single Legendre polynomials. Only then can their corresponding coefficients be compared in a meaningful least-squares fit.
To fit $B_f(W, x)$ by $B_g(W, x)$, we therefore expand both quantities in Legendre polynomials:
\begin{equation}
B_f(W, x) = \sum_{l=0}^{2N} F_l(W) P_l(x), \qquad
B_g(W, x) = \sum_{l=0}^{2M} G_l(W) P_l(x),
\end{equation}
where
\begin{equation}
F_l(W) = \sum_{i,j=0}^{N} a_i(W)\overline{a_j(W)}\, C_{ij}^{\ \ l},
\qquad
G_l(W) = \sum_{p,q=0}^{M} b_p(W)\overline{b_q(W)}\, C_{pq}^{\ \ l},
\end{equation}
and the Legendre product coefficients $C_{ij}^{\ \ l}$ are defined through
\begin{equation}
P_i(x)P_j(x) = \sum_l C_{ij}^{\ \ l} P_l(x).
\end{equation}

We minimize the standard Legendre least-squares error
\begin{equation}
\mathcal{N} = \|B_f(W, x) - B_g(W, x)\|^2
= \int_{-1}^{1} |B_f(W, x) - B_g(W, x)|^2 dx.
\end{equation}
Using orthogonality, this becomes
\begin{equation}
\mathcal{N}
=
\sum_{l=0}^{2M} \frac{2}{2l+1}|F_l(W)-G_l(W)|^2
+
\sum_{l=2M+1}^{2N} \frac{2}{2l+1}|F_l(W)|^2,
\end{equation}
since $G_l(W)=0$ for $l>2M$. The second sum is independent of $g(W,x)$, so minimizing $\mathcal{N}$ is equivalent to minimizing
the reduced functional
\begin{equation}
\mathcal{N}_{\mathrm{red}}
=
\sum_{l=0}^{2M} \frac{2}{2l+1}|F_l(W)-G_l(W)|^2.
\end{equation}
Thus the optimal truncated bilinear fit depends on $f(W,x)$ only through $(F_0,F_1,\ldots,F_{2M})$.

The parameters $b_0,\ldots,b_M$ do not arise from independent projections of the coefficients $a_n$. Instead, they are
obtained from a coupled nonlinear optimization problem because
\begin{equation}
G_l(W)=\sum_{p,q=0}^{M} b_p(W)\overline{b_q(W)}\,C_{pq}^{\ \ l},
\end{equation}
which is quadratic in the unknowns $b_m(W)$. Thus each $b_m(W)$ depends on the whole low-order vector
$(F_0,F_1,\ldots,F_{2M})$.

Therefore, the coefficients $b_m(W)$ depend on $f(W,x)$ only through the bilinear combinations
$a_i(W)\overline{a_j(W)}$ that contribute to $(F_0,F_1,\ldots,F_{2M})$. They do not, in general, depend separately or
linearly on the individual $a_n(W)$. Truncation in the bilinear problem therefore induces angular-momentum mixing
in the extracted coefficients: the fitted coefficients of $g(W,x)$ generally combine information from many coefficients
of $f(W,x)$, but only through the modes of $f(W,x)\overline{f(W,x)}$ up to order $2M$.

Although derived here for scalar scattering, the same algebraic point applies to photoproduction observables,
which are likewise bilinear in the underlying amplitudes.

One may, of course, always construct a truncated amplitude by simply discarding higher partial waves and retaining the lower coefficients unchanged. However, this does not in general coincide with the optimal fit of the bilinear observables. In the bilinear problem, the best-fit lower coefficients are determined by the coupled nonlinear minimization and therefore require reshuffling of the lower-order partial waves.

\subsection{Toy model: connection between the fitted and original amplitudes}

As it is difficult to visualize directly what is happening, we construct a simplified but analytically solvable model
in which all steps are transparent.

Consider the order-2 amplitude
\begin{equation}
f(x)=a_0P_0(x)+a_1P_1(x)+a_2P_2(x), \qquad a_0,a_1,a_2\in\mathbb{C},
\end{equation}
and approximate its Hermitian bilinear
\begin{equation}
B_f(x)=f(x)\overline{f(x)}
\end{equation}
by the order-1 Hermitian bilinear
\begin{equation}
B_u(x)=u(x)\overline{u(x)}, \qquad
u(x)=\alpha_0P_0(x)+\alpha_1P_1(x), \qquad \alpha_0,\alpha_1\in\mathbb{C}.
\end{equation}

Our problem is not to determine which coefficients $\alpha_0$ and $\alpha_1$ best represent the truncated amplitude itself. For the amplitude alone, orthogonality makes the answer trivial: the lower coefficients are simply retained. The actual problem is different. We want to determine which coefficients $\alpha_0$ and $\alpha_1$ in the bilinear $B_u(x)$ best represent the bilinear $B_f(x)$. To do this, both bilinears must first be transformed into the orthogonal basis of single Legendre polynomials, after which the corresponding coefficients can be compared.

The exact bilinear expansion $B_f(x)$ must therefore be transformed into its orthogonal form
\begin{equation}
B_f(x)=F_0P_0(x)+F_1P_1(x)+F_2P_2(x)+F_3P_3(x)+F_4P_4(x),
\end{equation}
where its coefficients are given by
\begin{equation}
F_0 = |a_0|^2+\frac{1}{3}|a_1|^2+\frac{1}{5}|a_2|^2,
\qquad
F_1 = 2\,\mathrm{Re}(a_0\overline{a_1})+\frac{4}{5}\,\mathrm{Re}(a_1\overline{a_2}),
\qquad
F_2 = 2\,\mathrm{Re}(a_0\overline{a_2})+\frac{2}{3}|a_1|^2+\frac{2}{7}|a_2|^2.
\end{equation}
These coefficients arise because products of Legendre polynomials are re-expanded as sums of single Legendre polynomials.

The same must be done for the lower-order bilinear $B_u(x)$:
\begin{equation}
B_u(x)=G_0P_0(x)+G_1P_1(x)+G_2P_2(x),
\end{equation}
where
\begin{equation}
G_0 = |\alpha_0|^2+\frac{1}{3}|\alpha_1|^2,
\qquad
G_1 = 2\,\mathrm{Re}(\alpha_0\overline{\alpha_1}),
\qquad
G_2 = \frac{2}{3}|\alpha_1|^2.
\end{equation}

If the order-1 bilinear ansatz reproduces the first three Legendre coefficients exactly, one has
\begin{equation}
G_0=F_0, \qquad G_1=F_1, \qquad G_2=F_2.
\end{equation}
This immediately yields the direct relations
\begin{equation}
|\alpha_1|^2 = \frac{3}{2}F_2,
\qquad
|\alpha_0|^2 = F_0-\frac{1}{2}F_2,
\qquad
\mathrm{Re}(\alpha_0\overline{\alpha_1})=\frac{1}{2}F_1.
\end{equation}
Substituting the expressions for $F_0$, $F_1$, $F_2$ gives the explicit connection to the original coefficients:
\begin{equation}
|\alpha_1|^2 = 3\,\mathrm{Re}(a_0\overline{a_2})+|a_1|^2+\frac{3}{7}|a_2|^2,
\end{equation}
\begin{equation}
|\alpha_0|^2 = |a_0|^2-\mathrm{Re}(a_0\overline{a_2})+\frac{2}{35}|a_2|^2,
\end{equation}
\begin{equation}
\mathrm{Re}(\alpha_0\overline{\alpha_1})=\mathrm{Re}(a_0\overline{a_1})+\frac{2}{5}\mathrm{Re}(a_1\overline{a_2}).
\end{equation}
These three equations constrain the fitted order-1 bilinear amplitude up to the usual phase ambiguities.

The important point is that the fitted lower-order quantities depend on combinations of all three original coefficients
$a_0$, $a_1$, and $a_2$. Therefore, if the original coefficients contain pole contributions in $W$, then the fitted
low-order quantities inherit mixed dependence on those pole-bearing terms. In that precise sense, truncation produces
mixing of resonance-carrying contributions already at the level of the fitted bilinear problem. The present toy model
does not attempt a separate analysis of the analytically continued poles of the reconstructed truncated amplitudes; its
purpose is to establish the mixing mechanism itself.

The above relations hold if the first three Legendre coefficients can be matched exactly by the order-1 Hermitian
bilinear ansatz. In general, this need not be the case. If no exact solution exists, the fitted coefficients $\alpha_0$ and
$\alpha_1$ are defined instead by the minimization of the reduced least-squares functional
\begin{equation}
\mathcal{N}_{\mathrm{red}}
=
2(F_0-G_0)^2+\frac{2}{3}(F_1-G_1)^2+\frac{2}{5}(F_2-G_2)^2.
\end{equation}
The explicit relations above should therefore be understood as the special case in which the low-order coefficients are
reproduced exactly.

It is nevertheless clear that $\alpha_0$ and $\alpha_1$ each depend on all three order-2 amplitude coefficients. The analytic
structure of the initial coefficients is therefore redistributed throughout the lower-order solution. Consequently, even if
one starts with distinct resonant contributions in the $S$, $P$, and $D$ partial waves, characterized by $L=0,1,2$, the
lower-order bilinear fit represents a superposition of these contributions.

\section{The problem of H\"ohler's clustering}

The phenomenon addressed in this paper was stated very explicitly by H\"ohler in his analysis of $\pi N$ partial waves~\cite{Hoehler1983}. Ordering the resonances according to the real part of their pole positions, he observed that several resonances belonging to different partial waves appear to cluster within the errors around a few common complex energies. As a concrete example, he noted that among the nucleon resonances there are sizable groups with pole positions near $(1670-50i)\,\mathrm{MeV}$ and $(2120-170i)\,\mathrm{MeV}$~\cite{Hoehler1983}. This is surprising because the poles in question belong to partial waves with different angular momenta and parities.

H\"ohler then raised the question whether the notion of resonances with well-defined quantum numbers is really the appropriate description of the phenomenon at intermediate energies. In particular, he suggested that $\pi N$ scattering might be describable by assuming a few complex poles of the invariant amplitudes instead of poles in many partial waves at slightly different positions~\cite{Hoehler1983}. In modern language, this amounts to asking whether the observed clustering reflects a genuine dynamical degeneracy of the exact amplitude, or whether it is produced effectively by the procedure used to extract poles from partial-wave analyses.

The historical observation must also be viewed in the light of present compilations. In the current PDG review, resonance properties are drawn from a broader set of hadronic and photoproduction analyses, and only the more firmly established states are retained in the summary tables~\cite{PDG2024}. As a result, the clustering pattern emphasized by H\"ohler is not displayed today in as direct a form as in the older KH-based compilations. At the same time, the PDG stresses that reliable resonance claims require a sufficiently complete partial-wave basis and explicit tests of the effect of higher partial waves~\cite{PDG2024}, which points directly to the relevance of truncation issues.

There are therefore two logically distinct possibilities. Either the clustering reflects a genuine dynamical feature of the exact amplitude, or it is an effective phenomenon generated by the extraction of truncated partial-wave information from bilinear observables. The present paper is concerned with the second possibility. Rather than attempting a general analysis of all possible dynamical origins of clustering, we focus on the truncation mechanism developed in the previous sections and examine how it can generate cross-wave correlations in the extracted resonance content.
\section{Possible connection to H\"ohler's clustering}

We can now return to the question posed in Sec.~3. The mechanism established in Sec.~2 shows that, in a truncated
bilinear fit, the extracted partial waves are not direct projections of the exact ones, but effective mixtures of pole-bearing
contributions from several angular-momentum sectors. From this point of view, it is natural to ask whether
H\"ohler-type clustering may be one phenomenological manifestation of the same effect.

In practical analyses one does not determine the amplitudes directly, but fits observables that are bilinear in them.
Once such observables are treated with a truncated ansatz, the extracted lower-order partial waves are obtained from
a coupled nonlinear problem and need not correspond to pure lower partial waves of the full non-truncated solution.
Instead, they carry mixed pole-bearing content.

This mixing is not limited to omitted higher partial waves. The nonlinear bilinear fit also reshuffles the pole-bearing
content already present in the nominally retained lower sector. Consequently, the partial waves obtained at a given
truncation order should not, in general, be interpreted as the exact lower partial waves with small corrections from
missing higher ones. Rather, they are effective partial waves in which lower and higher exact contributions are
recombined.

From this point of view, H\"ohler's clustering acquires a natural interpretation. If several fitted partial waves inherit
contributions from overlapping sets of exact amplitudes, then the resonance content extracted from them can exhibit
cross-wave correlations and may, in realistic truncated analyses, appear in groups with similar complex energies.
In this sense, clustering may arise as an effective property of the extraction procedure itself.

This may also help explain why H\"ohler-type clustering is much less visually obvious in present-day PDG summaries
than in the older phenomenological compilations: modern resonance listings are filtered through broader databases,
mixed extraction strategies, and stricter criteria on partial-wave completeness and robustness~\cite{PDG2024}. We do
not claim that truncation is the unique origin of H\"ohler's clustering. The claim is instead that it provides a natural
and unavoidable mechanism that can contribute to the type of cross-wave correlations emphasized by H\"ohler.
\section{Discussion and conclusions}

We have shown that truncation in a bilinear fitting problem is qualitatively different from a simple restriction of
the partial-wave expansion. Since the observables used to infer partial-wave content are bilinear in the amplitudes,
the extracted lower partial waves are determined by a coupled nonlinear set of equations rather than by direct
projection of the exact ones. As a result, the pole-bearing content of the exact infinite partial-wave decomposition is
redistributed in the truncated solution.

The central consequence is that the extracted partial waves should not, in general, be interpreted as carrying a
unique angular-momentum assignment in the same sense as the exact partial waves of the infinite problem. They are
truncation-dependent effective mixtures of contributions originating from several angular-momentum sectors. The toy
model makes this mechanism explicit: even when the fitted ansatz retains only lower partial waves, the reconstructed
lower-order quantities depend on bilinear combinations involving higher-order pieces of the original amplitude.

This provides a concrete mechanism for angular-momentum mixing in realistic truncated analyses. More importantly,
it clarifies the status of resonances extracted from TPWA. In general, the resonance-related quantities obtained from
a truncated analysis should not be identified with direct projections of the corresponding quantities of the exact
non-truncated problem. They are effective structures generated by the restricted bilinear fit.

From this point of view, H\"ohler's clustering can be understood as one possible phenomenological manifestation
of a more general effect. If the extracted partial waves inherit overlapping pole-bearing contributions from several
exact angular-momentum sectors, then the inferred resonance spectrum can exhibit cross-wave correlations and may,
in realistic truncated analyses, appear in groups with similar complex energies.

We do not claim that truncation is the unique origin of H\"ohler's clustering. The claim is instead that it provides a
natural and unavoidable mechanism that can contribute to the type of cross-wave correlations emphasized by H\"ohler.
The practical message is therefore clear: TPWA remains an extremely valuable and often indispensable tool, but
the resonances it extracts must be interpreted with appropriate caution. A truncated analysis does not, in general,
reveal the true resonance content of the full problem in a direct way. It yields instead a truncation-dependent effective
representation of that content.

\section*{Acknowledgments}
I would like to express my gratitude to Ron Workman for raising the issue of H\"ohler's clustering. I acknowledge having used the Overleaf AI language model to refine the wording of the text. I also acknowledge having used the large language model ChatGPT 5.4 to verify the formal correctness of the presentation and to evaluate the completeness of the cited material. I accept full responsibility for the validity of the physical assumptions and for all conclusions drawn in the text.

\end{document}